\documentclass[aps,prl,twocolumn,showpacs,preprintnumbers,amsmath,amssymb]{revtex4-1}

 \def\frac#1#2{{#1\over #2}}

\def\be{\begin{equation}}
\def\ee{\end{equation}}
\def\ba{\begin{eqnarray}}
\def\ea{\end{eqnarray}}

\def\({\left(}
\def\){\right)}

\def\[{$$}
\def\]{$$}
\def\({$}
\def\){$}

\usepackage{color}
\usepackage{graphicx}
\usepackage{dcolumn}
\usepackage{bm}
\usepackage{ulem}
\usepackage{appendix}
\usepackage{tikz}
\usepackage{amsmath}
\usepackage[thicklines]{cancel}
\usetikzlibrary{decorations.pathreplacing}

\begin{document}

\title{Neural Ordinary Differential Equations for Mapping the Magnetic QCD Phase Diagram via Holography}

\author{Rong-Gen Cai$^{*b,c,f}$, Song He$^{*b,a,e}$, Li Li$^{*c,d,f}$, Hong-An Zeng$^{*a}$}
\affiliation{$^a$Center for Theoretical Physics and College of Physics, Jilin University,\\ Changchun 130012, People's Republic of China}
\affiliation{$^{b}$Institute of Fundamental Physics and Quantum Technology, \& School of Physical Science and Technology,
Ningbo University, Ningbo, Zhejiang 315211, China}
\affiliation{$^{c}$Institute of Theoretical Physics, Chinese Academy of Sciences, Beijing 100190, China}
\affiliation{$^{d}$School of Fundamental Physics and Mathematical Sciences, Hangzhou Institute for Advanced Study, UCAS, Hangzhou 310024, China}
\affiliation{$^{e}$Max Planck Institute for Gravitational Physics (Albert Einstein Institute),\\ Golm 14476, Germany}
\affiliation{$^{f}$School of Physical Sciences, University of Chinese Academy of Sciences, Beijing 100049, China.}

\email{cairg@itp.ac.cn,\\hesong@nbu.edu.cn,\\liliphy@itp.ac.cn,\\zengha20@mails.jlu.edu.cn}

\begin{abstract}
The QCD phase diagram is crucial for understanding strongly interacting matter under extreme conditions, with major implications for cosmology, neutron stars, and heavy-ion collisions. We present a novel holographic QCD model utilizing neural ordinary differential equations (ODEs) to map the QCD phase diagram under magnetic field $B$, baryon chemical potential $\mu_B$, and temperature $T$. By solving the inverse problem of constructing a gravitational theory from Lattice QCD data, we reveal an unprecedentedly rich phase structure at finite $B$, including multiple critical endpoints (CEPs) in strong magnetic fields. Specifically, for {$B = 1.618 \, \mathrm{GeV}^2=2.592 \times 10^{19}$ Gauss}, we identify two distinct CEPs at $(T_C = 87.3 \, \mathrm{MeV}, \, \mu_C = 115.9 \, \mathrm{MeV})$ and $(T_C = 78.9 \, \mathrm{MeV}, \, \mu_C = 244.0 \, \mathrm{MeV})$. Notably, the critical exponents vary depending on the CEP's location, and the conventional scaling relations can be violated in the presence of strong magnetic fields. These findings significantly advance our understanding of the QCD phase structure and provide concrete predictions for experimental validation at upcoming facilities such as FAIR, JPARC-HI, and NICA.

\end{abstract}


\maketitle
\section{I. Introduction}
Quantum Chromodynamics (QCD), the theory of strong interactions, governs quark and gluon behavior under extreme conditions, underpinning high-energy physics, astrophysics, and cosmology. The QCD phase diagram, which maps transitions between the deconfined quark-gluon plasma and hadronic matter, is central to understanding phenomena in non-central heavy-ion collisions, neutron stars, and the early universe~\cite{Braun-Munzinger:2008szb, Philipsen:2012nu, Gupta:2011wh}. The global structure of the QCD phase diagram, spanning temperature ($T$), baryon chemical potential ($\mu_B$), and magnetic field ($B$), remains poorly understood~\cite{Andersen:2014xxa}. At low $\mu_B$ and $B=0$, QCD matter undergoes a smooth crossover, while at high $\mu_B$, a first-order phase transition occurs. Locating the critical endpoint (CEP) separating these regimes is a primary goal of heavy-ion collision experiments~\cite{Aoki:2006we, HotQCD:2018pds}. Magnetic fields, generated in non-central heavy-ion collisions and present in astrophysical objects like magnetars, significantly influence the QCD equation of state. Strong magnetic fields modify the critical temperature, induce novel phases, and play a key role in shaping the QCD phase structure~\cite{Andersen:2014xxa}, making $B$ an essential factor in searching for CEP experimentally~\cite{STAR:2020tga,STAR:2021fge,STAR:2022hbp}. Exploring the magnetic QCD phase structure remains a significant challenge.

With substantial progress, current non-perturbative approaches to studying the QCD phase diagram still face significant challenges. As a first-principle approach, Lattice QCD encounters the well-known \textit{sign problem} at finite baryon density, making numerical simulations at high $\mu_B$ computationally prohibitive~\cite{Philipsen:2012nu}. Moreover, simulating high-magnetic-field configurations requires substantial computational resources, limiting exploration of the QCD phase structure in these regions~\cite{Andersen:2014xxa}. These limitations significantly hinder our understanding of the interplay among $T$, $\mu_B$, and $B$, particularly in the regions of the phase diagram where experimental observations are most challenging.

As a competitive approach, holography offers a computationally efficient framework to study non-perturbative QCD systems by mapping them to classical gravity~\cite{Demircik:2021zll, Chen:2022goa, Rougemont:2023gfz}. The holographic framework has particularly effectively captured QCD dynamics at finite temperatures and densities~\cite{Gubser:2008yx, DeWolfe:2010he, DeWolfe:2011ts}. Several refined models, only capturing corner of the QCD phase diagram, still fall short of quantitatively describing the full set of lattice QCD data, even for QCD equations of state~\cite{Cai:2012xh, He:2013qq, Alho:2013hsa, Critelli:2017oub, Knaute:2017opk, Gursoy:2017wzz, Yang:2020hun, Grefa:2021qvt}.  
To construct a holographic model consistent with QCD data, one must systematically explore various actions and scan the parameter space—a task akin to finding a needle in a haystack. Manual parameter tuning to fit lattice QCD data, known as the \textit{inverse problem}, is computationally inefficient and often fails to capture the system's complexity, particularly in the multi-dimensional parameter space of $T$, $\mu_B$, and $B$~\cite{Finazzo:2016mhm, Rougemont:2023gfz, Gursoy:2017wzz}.

Neural ordinary differential equations (neural ODEs) excel in modeling complex, continuous dynamic systems by optimizing over infinite-dimensional parameter spaces with adaptive precision. In this work, we leverage neural ODEs in holography to address these challenges, offering a data-driven mechanism that surpasses traditional machine learning methods~\cite{Hashimoto:2018bnb, Akutagawa:2020yeo, Hashimoto:2020jug, Hashimoto:2022eij, Chen:2024ckb}. This innovative approach captures the intricate dependencies of thermodynamic quantities on $T$, $\mu_B$, and $B$ with unprecedented accuracy and flexibility.

These advantages allow our approach to yield a concrete global structure of the phase diagram and uncover unprecedented features of the QCD phase structure, such as multiple CEPs and their associated critical behaviors, which were previously unobserved. In particular, for sufficiently large $B = 1.618 \, \mathrm{GeV}^2=2.592 \times 10^{19}$ Gauss, the $T$-$\mu_B$ plane exhibits two CEPs, a feature exceeding prior expectations. Furthermore, we analyze the critical behavior near the CEPs and find that the critical exponents vary depending on the location of the CEP. Notably, the conventional scaling relations can be violated in the presence of strong magnetic fields. This innovative combination of machine learning and holography significantly advances our understanding of the QCD phase structure and provides new avenues for better agreement with experimental and lattice QCD results.

This work is organized as follows. We introduce the holographic model in Section II. We then describe our method for fixing the bulk theory using machine
learning in Section III. The full QCD phase diagram and its key features are summarized in Sections IV and V.
We conclude and provide future directions in Section VI. In Appendix A, we additionally present details for solving the equations of motion and the definitions of thermodynamic quantities. A detailed algorithmic breakdown of the neural ODEs is included in Appendix B.

\begin{figure}[hbt!]
	\centering
     \includegraphics[width=0.5\textwidth]{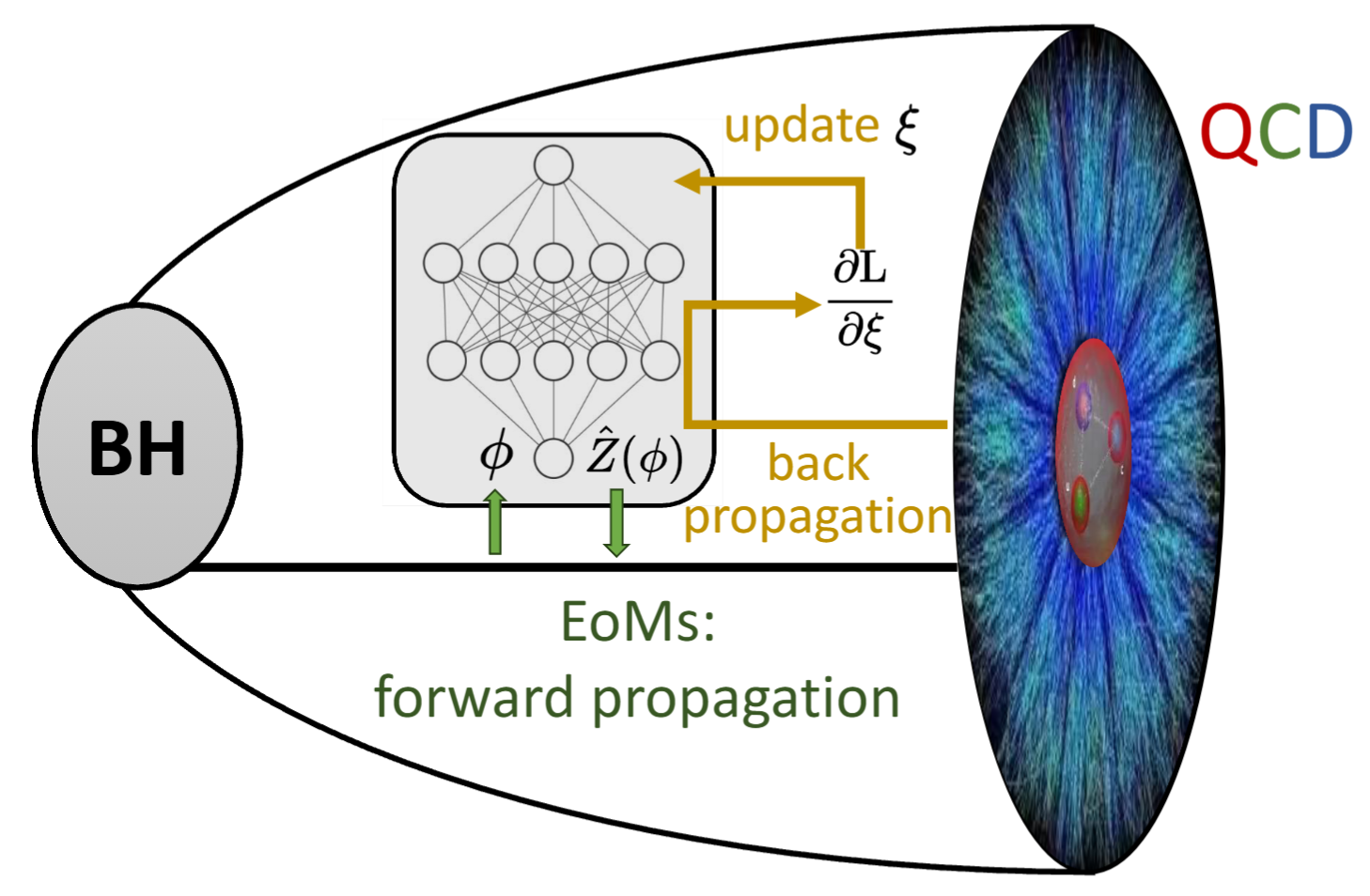}
	\caption{Schematic of the neural ODE approach for solving the inverse problem in QCD phase diagram analysis. Initial conditions from the black hole (BH) horizon are connected to QCD boundary data via EoMs. The neural network initialized as a trial function for the coupling $\hat{Z}(\phi)$ is optimized via backpropagation to minimize the loss function $\text{L}$, representing the difference from lattice QCD data. Parameters $\xi$ are optimized using gradient descent. Technical details are provided in Appendix B.}\label{fig:ODE}
\end{figure}

\section{II. Holographic model}\label{sec:model}
To capture essential QCD dynamics at finite magnetic field, temperature, and baryon chemical potential, we employ a holographic framework based on the five-dimensional gravitational theory:
\begin{eqnarray}\label{action}
S=\frac{1}{2\kappa_N^2}\int d^{5}x \sqrt{-g} \Big[\mathcal{R}-\frac{1}{2}\nabla_\mu \phi \nabla^\mu \phi \nonumber \\
-\frac{Z(\phi)}{4}F_{\mu\nu}F^{\mu\nu}-\frac{\hat{Z}(\phi)}{4}\hat{F}_{\mu\nu}\hat{F}^{\mu\nu}-V(\phi)\Big]\,,
\end{eqnarray}
where $\kappa_N^2$ is the effective Newton constant. The metric $g_{\mu\nu}$ characterizes spacetime geometry, and the real scalar field $\phi$ accounts for conformal symmetry breaking. The Maxwell field $A_\mu$ with $F_{\mu\nu}=\partial_\mu A_\nu-\partial_\nu A_\mu$ introduces a finite baryon number density, while the magnetic field $B$ is described by another Maxwell field $\hat{A}_\mu$ with $\hat{F}_{\mu\nu}=\partial_\mu \hat{A}_\nu-\partial_\nu \hat{A}_\mu$. The functions $Z(\phi)$, $\hat{Z}(\phi)$, and $V(\phi)$ that encode the non-perturbative features of our system are calibrated against lattice QCD data. The form of $V(\phi)$ and $Z(\phi)$ is taken from~\cite{Cai:2022omk}.
\begin{equation}\label{VZapp}
\begin{aligned}
V(\phi)&=-12\cosh[c_1\phi]+(6 c_1^2-\frac{3}{2})\phi^{2}+{c_2\phi^{6}}\,,\\
Z(\phi) &=\frac{1}{1+c_{3}} {\text{sech}[c_4\phi^3]}+\frac{c_{3}}{1+c_{3}}e^{-c_{5} \phi}\,.
\end{aligned}
\end{equation}
Note that the effective number of model parameters cannot be reduced by coordinate or gauge transformations. Coordinate transformations simplify the metric but leave \( V(\phi) \) and \( Z(\phi) \) unchanged, where the parameters \( c_1, c_2, c_3, c_4, c_5 \) govern the scalar field dynamics (see Appendix A). Similarly, gauge transformations simplify the gauge fields without affecting the parameters in \( V(\phi) \) or \( Z(\phi) \), which remain independent of the gauge choice. The comparison of the holographic model~\eqref{VZapp} to experimental observations can be found in~\cite{Li:2023mpv,He:2023ado}.

By varying the action~\eqref{action}, the equations of motion (EoMs) can be obtained:
 \begin{equation}
 \begin{aligned}
&\nabla_{\mu} \nabla^{\mu} \phi-\frac{\partial_{\phi} Z}{4} F_{\mu \nu} F^{\mu \nu}-\frac{\partial_{\phi} \hat{Z}}{4} \hat{F}_{\mu \nu} \hat{F}^{\mu \nu}-\partial_{\phi} V=0\,, \\
&\quad \quad \quad \quad \quad  \quad \quad \quad \quad \quad \quad \quad \quad  \quad \quad \nabla^{\nu}(Z F_{\nu \mu})=0\,, \\
&\quad \quad \quad  \quad \quad \quad \quad \quad \quad \quad \quad \quad \quad \quad \quad \nabla^{\nu}(\hat{Z} \hat{F}_{\nu \mu})=0\,, \\
&R_{\mu \nu}-\frac{1}{2} R g_{\mu \nu}=\frac{1}{2} \nabla_{\mu} \phi \nabla_{\nu} \phi+\frac{Z}{2} F_{\mu \rho} {F_{\nu}}^{\rho}+\frac{\hat{Z}}{2} \hat{F}_{\mu \rho}{\hat{F}_\nu}^{\rho}\\
&+\frac{1}{2}\left(-\frac{1}{2} \nabla_{\mu} \phi \nabla^{\mu} \phi-\frac{Z}{4} F_{\mu \nu} F^{\mu \nu}-\frac{\hat{Z}}{4} \hat{F}_{\mu \nu} \hat{F}^{\mu \nu}-V\right) g_{\mu \nu}\,.
 \end{aligned}\label{fig:EOM} 
\end{equation}

The bulk black hole solutions in Anti-de Sitter spacetime (AdS) are
\begin{equation}\begin{aligned}\label{EOM}
ds^2 = -f(r) e^{-\eta(r)} dt^2 + \frac{dr^2}{f(r)} + r^2 (dx^2 + dy^2 + g(r) dz^2)\,, \\
\phi = \phi(r), \; A_\mu dx^\mu=A(r)dt,\;  \hat{A}_\mu dx^\mu=\frac{B}{2} (x dy - y dx)\,,
\end{aligned}
\end{equation}
where $r$ is the holographic radial coordinate with the AdS boundary located at $r\rightarrow\infty$. The source term $\phi_s\equiv\lim_{r\rightarrow\infty}r\phi(r)$ essentially breaks the conformal symmetry and plays the role of the energy scale. Substituting the above ansatz into~\eqref{fig:EOM} yields the bulk equations that will be solved numerically.

The blackening function $f(r)$ is vanishing at the event horizon $r=r_h$. The temperature and entropy density of the QCD system are given by the famous Hawking temperature and Bekenstein-Hawking entropy of a black hole:
\begin{equation}
T=\frac{1}{4\pi}f'(r_h)e^{-\eta(r_h)/2},\quad s=\frac{2\pi}{\kappa_N^2} r_h^3\,,
\end{equation}
evaluated at the black hole event horizon. Note that the magnetic field $B$ breaks Lorentz invariance along the $z$-axis, leading to anisotropic pressure. Solving the bulk EoMs allows us to extract thermodynamic quantities via the holographic renormalization, such as the free energy density $\Omega$, energy density $\epsilon$, longitudinal pressure $p_z$, baryon density $n_B$, and magnetization $M$. Moreover, they are found to satisfy the expected thermodynamic laws:
\begin{equation}
\begin{split}
 T s&=\epsilon-\Omega-\mu_B n_B\,, \\ d\Omega&=-s dT-n_B d\mu_B-M dB\,. 
 \end{split}
\end{equation}
For further details, see Appendix A. This holographic approach provides a phenomenological description of QCD dynamics under extreme conditions. We have used units $c=\hbar =k_B=e=1$.
\begin{figure}[hbt!]
    \centering
    \includegraphics[width=0.94\linewidth]{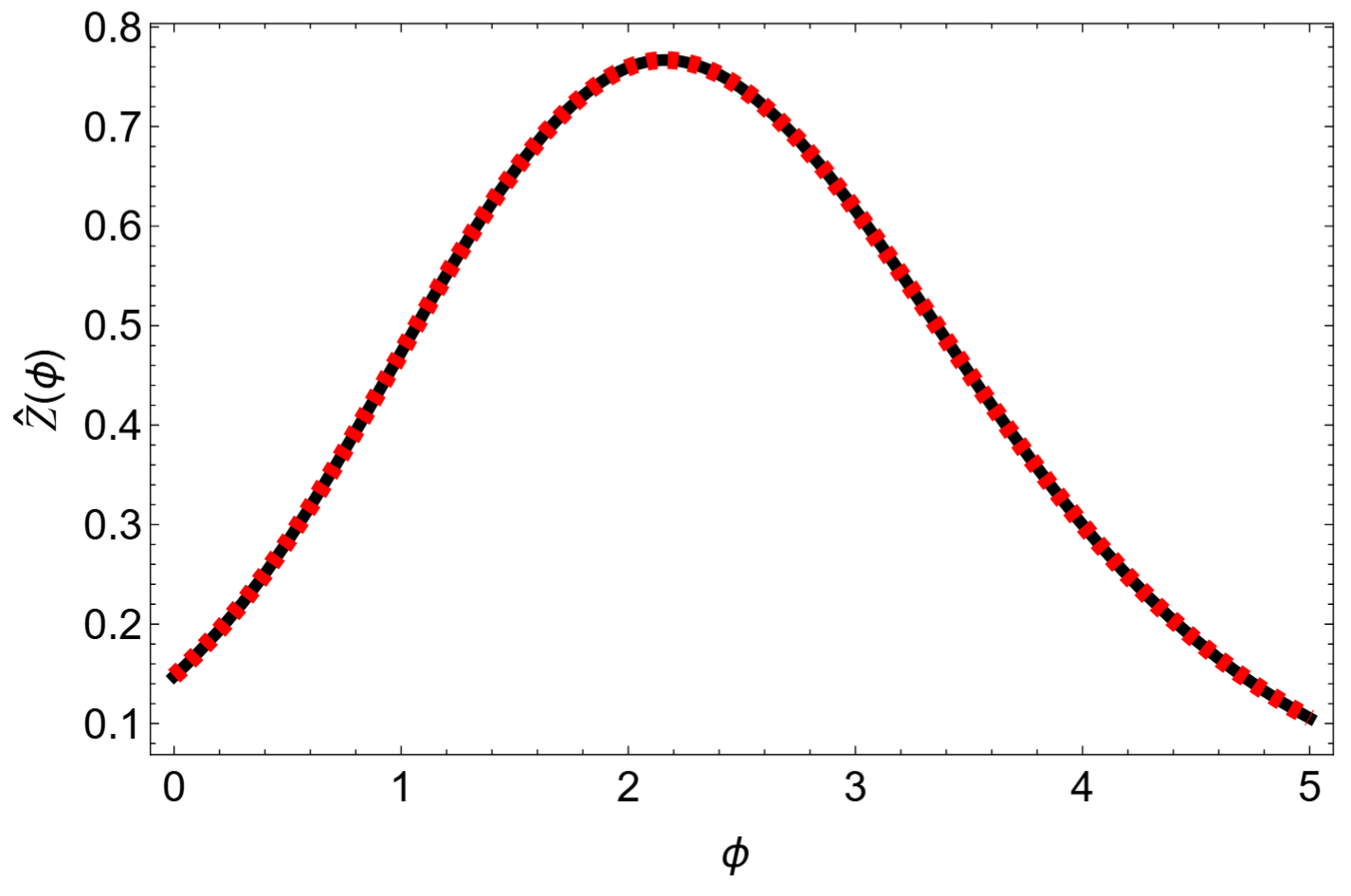}
    \caption{The magnetic coupling $\hat{Z}(\phi)$ as a function of $\phi$ from machine learning. The black solid curve is the one obtained from our neural ODEs architecture, and the red dotted one from the fitting function~\eqref{myfit} in Appendix B.}
    \label{fig:AIZphi}
\end{figure}
\begin{figure*}[hbt]
	\centering    
    \includegraphics[width=0.9\textwidth]{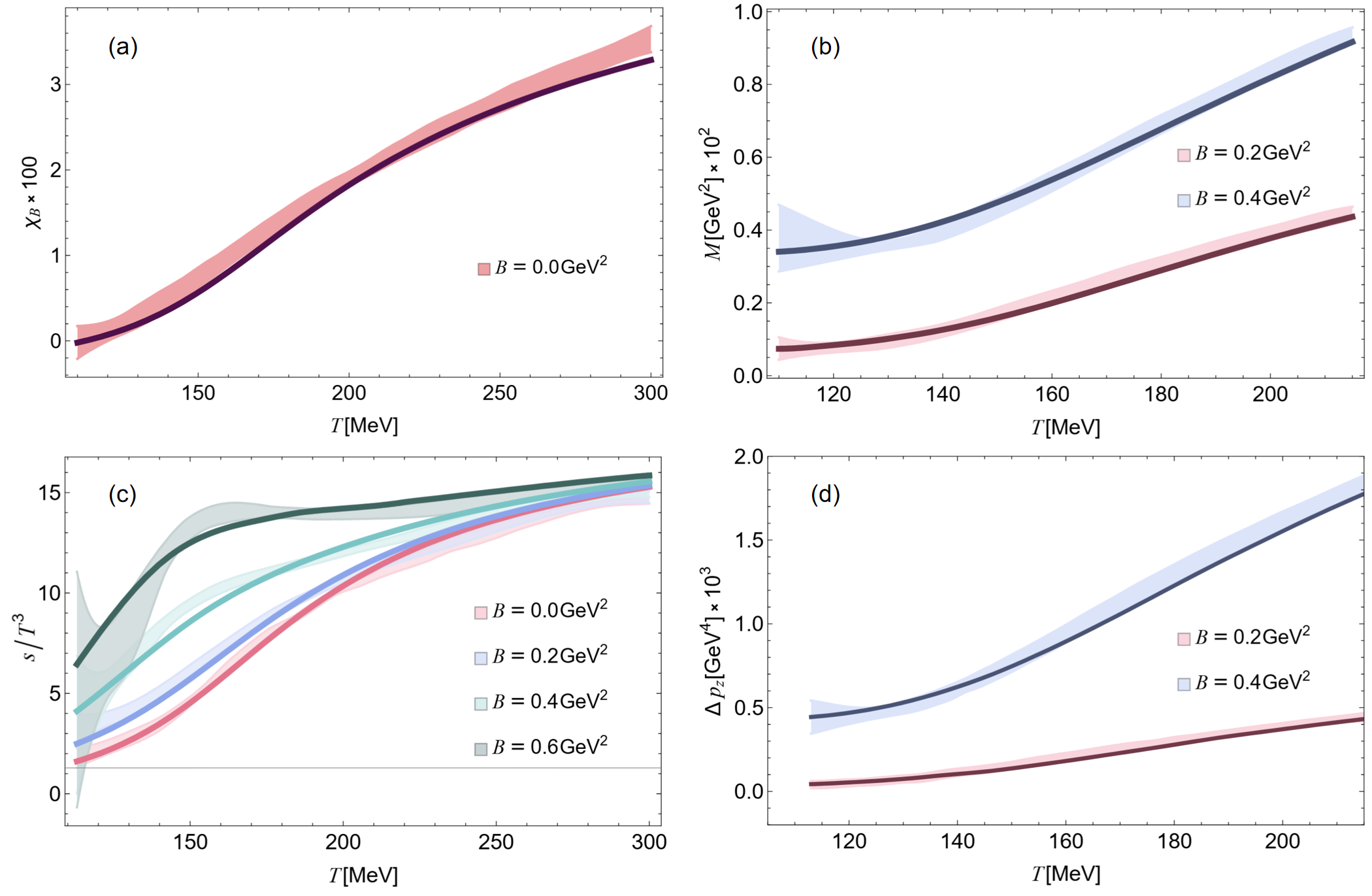} 
	\caption{Thermodynamic Quantities from Holographic QCD Model vs. Lattice Data. Temperature dependence of (a) magnetic susceptibility $\chi_B$, (b) magnetization $M$, (c) entropy density $s/T^3$, and (d) longitudinal pressure $\Delta p_z = p_z|_B - p_z|_{B=0}$ across magnetic fields. The shaded regions show lattice QCD estimates~\cite{Bali:2014kia}; the solid lines indicate model predictions. Here, $e = 1$, giving $B = 1 \, \text{GeV}^2 = 1.602 \times 10^{19}$ Gauss.}
	\label{fig:LQCDcomp}
\end{figure*}

\section{III. Neural ODEs}
In the absence of a first-principle method to determine the coupling functions in our bottom-up model, we constrain these functionals using available lattice QCD data—a challenge known as the inverse problem. As lattice QCD data at finite $B$ and $\mu_B$ become more abundant, manually tuning control parameters becomes impractical. Our neural ODE approach systematically explores the infinite-dimensional parameter space of the coupling functions, achieving higher precision and accuracy.

In practice, we impose boundary conditions at the ultraviolet (UV) boundary and the black hole event horizon. By solving the EoMs numerically, the neural ODE generates a trial equation of state, which is then iteratively optimized via backpropagation to reduce deviations from lattice QCD data. This process efficiently converges to an optimal magnetic coupling, providing a good agreement with lattice data and refining model predictions across unexplored magnetic fields and chemical potentials (see Fig.~\ref{fig:ODE} for illustration). A detailed algorithmic breakdown of this methodology is included in Appendix B. We construct a (2+1)-flavor holographic QCD (hQCD) model using this neural ODE approach to achieve a precise fit with lattice QCD results~\cite{Bali:2014kia}. A similar neural ODE-based method is applied to determine $Z(\phi)$ and $V(\phi)$ by matching to data at $B=0$~\cite{HotQCD:2014kol, Borsanyi:2021sxv}. The resulting functional forms are benchmarked rigorously and agree with recent lattice simulations and experimental data~\cite{Cai:2022omk, Li:2023mpv}. The profile for $\hat{Z}(\phi)$ from our neural ODEs is presented in Fig.~\ref{fig:AIZphi}. This strong non-monotonicity highlights the necessity for parameter tuning in machine learning.

Fig.~\ref{fig:LQCDcomp} presents our holographic predictions for four independent thermodynamic quantities: magnetic susceptibility $\chi^B$, magnetization $M$, entropy density $s$, and longitudinal pressure $p_z$, compared to lattice QCD data~\cite{Bali:2014kia}. We find good agreement across the available magnetic fields, supporting our holographic model. This is the first holographic model to achieve good agreement with lattice data for magnetic fields up to $B = 0.6 \text{ GeV}^2$. Further analysis is provided in Appendix B, showing consistency with lattice QCD results~\cite{Bali:2014kia}. While Fig. 3 demonstrates an overall reasonable agreement with the lattice data, especially for the entropy density, there are deviations from the central value, which can be as large as $2\sigma$, particularly at finite magnetic fields. These discrepancies are due to several factors, including the inherent limitations of the neural ODE approach and the simplifications made in the model. The neural ODE framework, while powerful for data-driven optimization, has certain limitations in capturing the full complexity of the system. This includes challenges related to the parameter space and the trade-off between model complexity and fitting accuracy. Further refinement of the model and optimization process could help reduce these deviations in future work.
\begin{figure}[hbt!]
	\centering
     \includegraphics[width=0.48\textwidth]{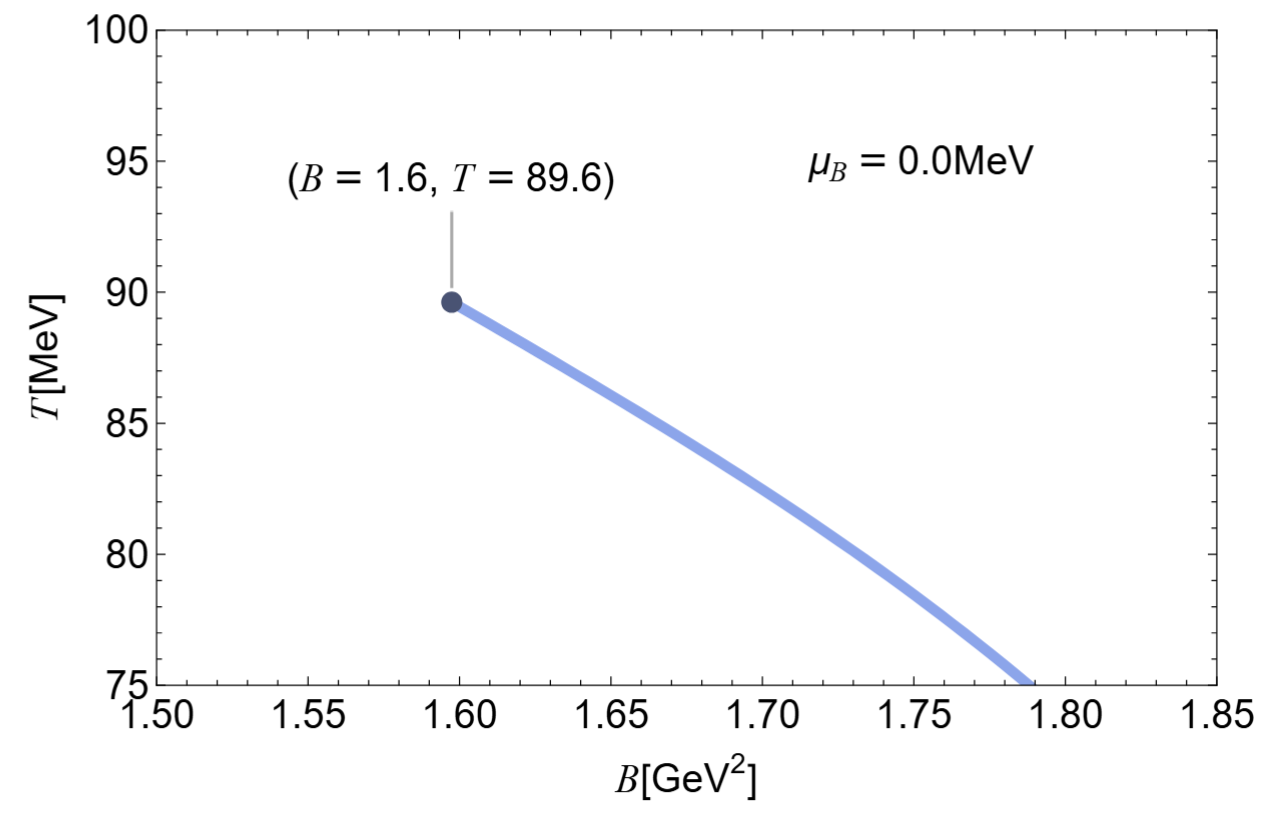}
	\caption{The phase diagram on the $B$-$T$ plane at vanishing $\mu_B$. The blue dot denotes the CEP, and the blue line corresponds to the first-order line.}\label{TB-mu-0}
\end{figure}

\section{IV. QCD phase diagram}
\begin{figure}[hbt!]
	\centering       \includegraphics[width=0.48\textwidth]{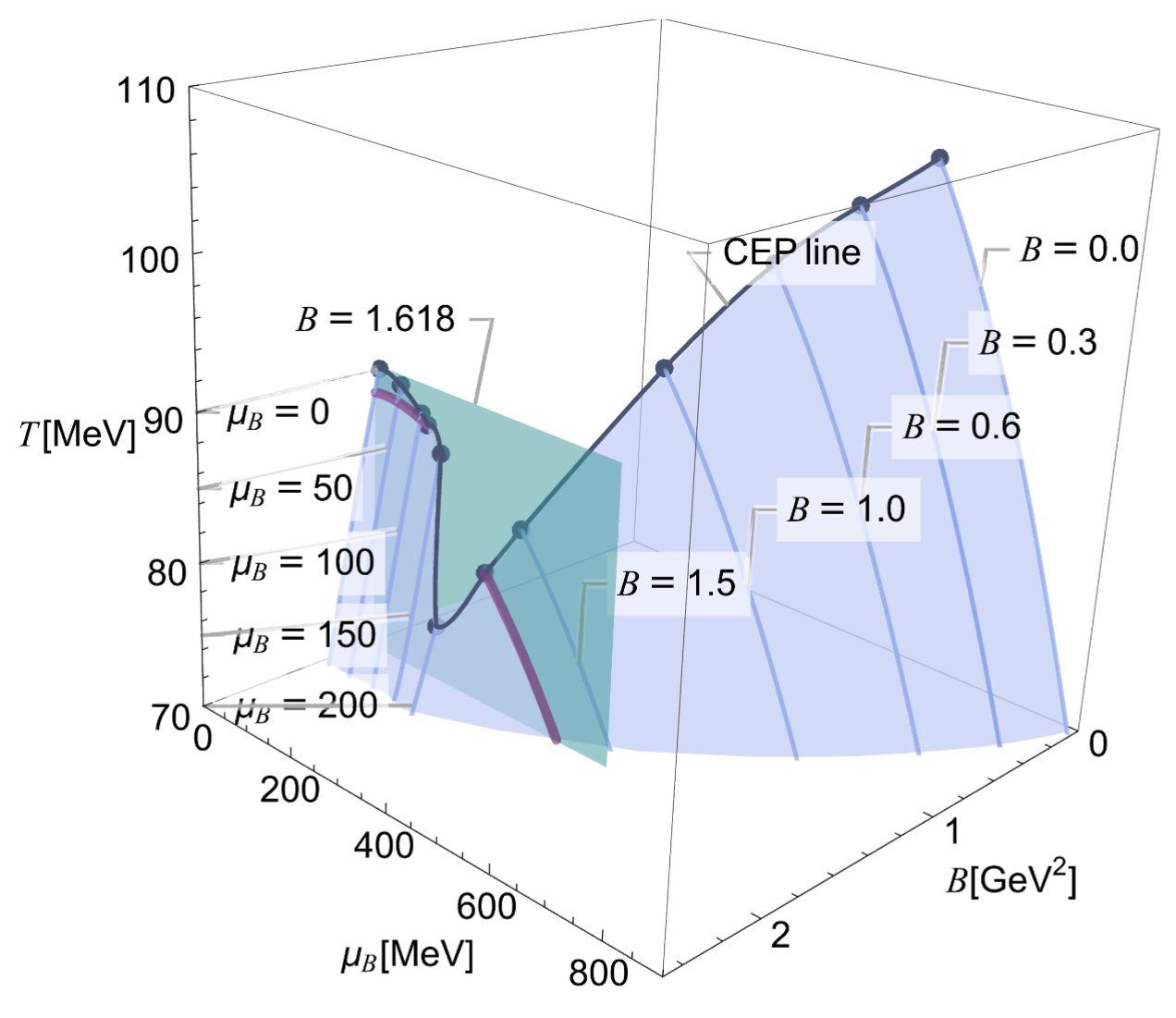}
        
	\caption{QCD phase diagram at finite magnetic field $B$. Phase structure in temperature $T$, baryon chemical potential $\mu_B$, and magnetic field $B$ from our holographic model. The light blue surface denotes the first-order transition boundary, separating the hadronic phase from the quark-gluon plasma. The dark blue line traces the CEP trajectory, marking where the first-order transition ends in a crossover.}
	\label{fig:lattice_comparison1}
\end{figure}

With the model fully established, we construct the QCD phase diagram at finite $B$, $T$, and $\mu_B$ by computing the free energy density $\Omega$. The full phase diagram is depicted in Fig.~\ref{fig:lattice_comparison1}. The light blue area denotes the first-order phase transition surface, dividing the quadrant into two parts: the high-temperature region corresponds to the quark-gluon plasma, while the low-temperature region corresponds to the hadron gas phase. The deep blue line in the diagram marks the location of CEP for various magnetic fields, where the first-order phase transition terminates and transitions into a smooth crossover at small chemical potentials. 
At $B=0$, the phase diagram was presented in Fig.~3 of~\cite{Cai:2022omk}, where the first-order transition line terminates at $(T_C = 105\,\text{MeV}, \mu_C = 555\,\text{MeV})$. Fig.~\ref{TB-mu-0} shows the phase structure in the $T$-$B$ plane at $\mu_B = 0$, revealing a line of first-order transitions ending at the CEP located at $(T_C = 89.6\,\text{MeV}, B = 1.6\,\text{GeV}^2)$, consistent with lattice QCD predictions~\cite{Cuteri:2023evl}.

\begin{figure}[hbt!]
	\centering       \includegraphics[width=0.46\textwidth]{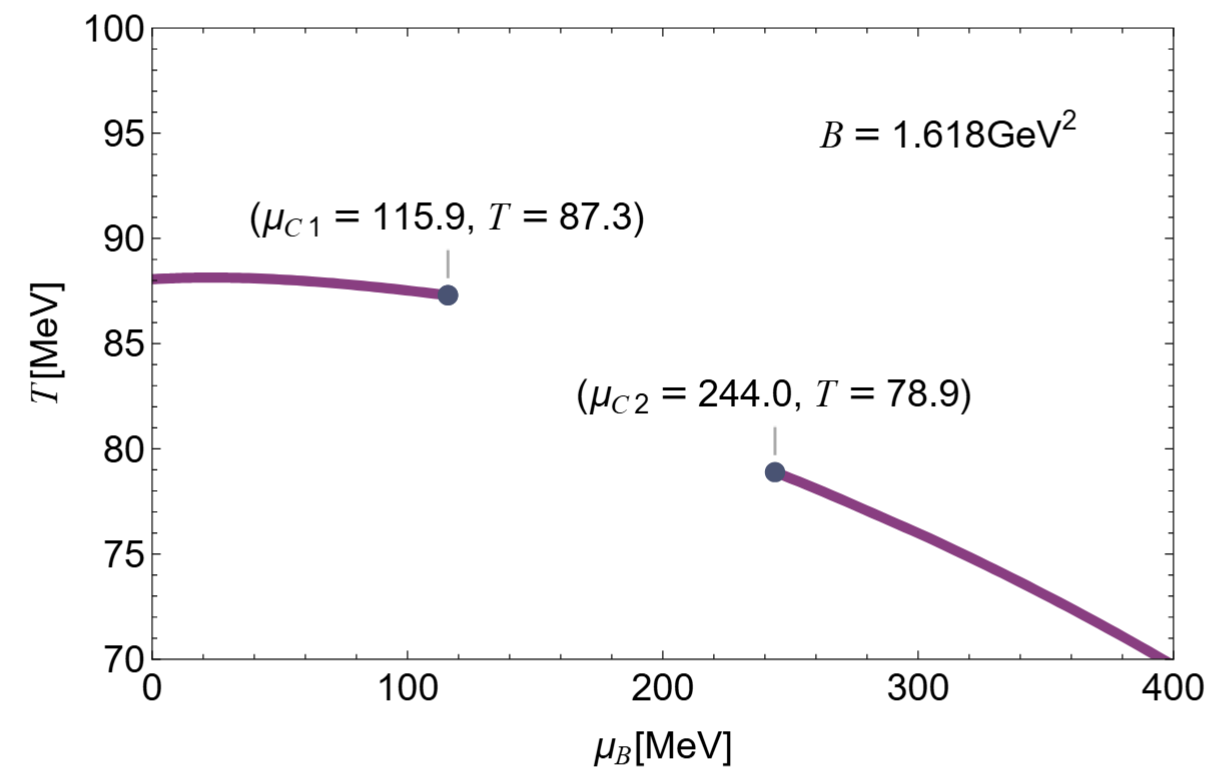}
    \caption{QCD phase diagram in the $T$-$\mu_B$ plane at $B = 1.618\, \mathrm{GeV}^2$. The purple line marks the first-order phase transition, ending at the first critical endpoint (CEP) at $\mu_{C1} = 115.9$ MeV, where the transition becomes a crossover. A second CEP at $\mu_{C2} = 224.0$ MeV suggests a more complex phase structure. These findings imply that strong magnetic fields induce rich QCD phase behavior, with significant implications for heavy-ion collisions.}
\label{muT-twoCEP}
\end{figure}

Fig.~\ref{fig:lattice_comparison1} highlights the following three key observations.
\begin{enumerate}
    \item As the magnetic field $B$ increases up to $B = 1.764 \, \text{GeV}^2$, the critical chemical potential $\mu_C$ at the CEP decreases, indicating that stronger magnetic fields shift the CEP to lower chemical potentials. This shows a significant effect of magnetic fields on the CEP location, which affects conditions in experimental settings such as heavy-ion collisions.
    
    \item The critical temperature $T_C$ at the CEP initially decreases with increasing $B$, reaching a minimum before increasing again. This turning point occurs around $T =80 \text{MeV}$, $B = 1.6 \sim 1.7 \text{GeV}^2$, and $\mu_B = 0.2 \sim 0.28 \text{GeV}$. This behavior suggests complicated effects in the presence of a background magnetic field. It could be related to the inverse magnetic catalysis and magnetic catalysis reported in the literature.
    
    \item Multiple CEPs emerge at sufficiently strong magnetic fields in the $T$-$\mu_B$ phase diagram, as shown in Fig.~\ref{muT-twoCEP}. A first-order phase transition occurs for $0 < \mu_B < \mu_{C1}$ and $\mu_B > \mu_{C2}$, while a crossover exists when $\mu_{C1} < \mu_B < \mu_{C2}$. As the magnetic field $B$ increases, the two CEPs, initially distinct, converge into a single point. Specifically, Fig.~\ref{muT-twoCEP} illustrates two CEPs identified at $(T_C, \mu_C) = (87.3 \, \mathrm{MeV}, 115.9 \, \mathrm{MeV})$ and $(T_C, \mu_C) = (78.9 \, \mathrm{MeV}, 244.0 \, \mathrm{MeV})$. This result holds significant implications for heavy-ion collision experiments, providing precise experimental markers for testing at FAIR, JPARC-HI, and NICA~\cite{Fukushima:2020yzx}.
\end{enumerate}
These key observations reveal a rich phase structure in a strong magnetic field and warrant further experimental verification.

\section{V. Critical exponents}
Beyond mapping the phase diagram, we examine critical behavior near the CEPs via critical exponents. These exponents describe how thermodynamic quantities, such as susceptibility and specific heat, diverge near critical points, typically following power-law scaling. These exponents help identify the CEP's universality class and provide insights into QCD transitions under extreme conditions.

Four critical exponents can be directly extracted from the phase diagram of Fig.~\ref{fig:lattice_comparison1}.
\begin{itemize}
    \item \textbf{ 
    Critical exponent \(\alpha\):} The exponent $\alpha$ quantifies the power-law behavior of specific heat near a CEP along the axis defined as approaching the CEP along the tangent of the first-order line:
    \[ 
    C_n = T \left(\frac{\partial s}{\partial T}\right)_{n_B,B} \sim |T - T_{\text{CEP}}|^{-\alpha}.
    \]
    \item \textbf{
    Critical exponent \(\beta\):} It characterizes the discontinuity of entropy density $s$ across the first-order line:
    \[ 
    \Delta s = s_> - s_< \sim (T_{\text{CEP}} - T)^\beta,
    \]
    where \(s_>\) and \(s_<\) represent the entropy densities in the high- and low-temperature phases, respectively.
    
    \item \textbf{
    Critical exponent \(\gamma\):} It represents the power-law behavior of baryon susceptibility with the temperature near the CEP along the first-order axis:
    \[ 
    \chi_2^{B} = \frac{1}{T^2}\left( \frac{\partial n_B }{\partial \mu_B} \right)_{T,B} \sim |T - T_{\text{CEP}}|^{-\gamma}.
    \]
    \item \textbf{Critical exponent \(\delta\) 
    }: The definition of $\delta$ relies on the power-law relationship between entropy and chemical potential with $T=T_{\text{CEP}}$ at the critical isotherm:
    \[ 
    s - s_{\text{CEP}} \sim |\mu_B - \mu_{B\text{CEP}}|^{1/\delta},
    \]
    where \(s_{\text{CEP}}\) is the entropy density at the CEP.
\end{itemize}
\begin{figure*}[hbt]
	\centering    
    \includegraphics[width=0.85\textwidth]{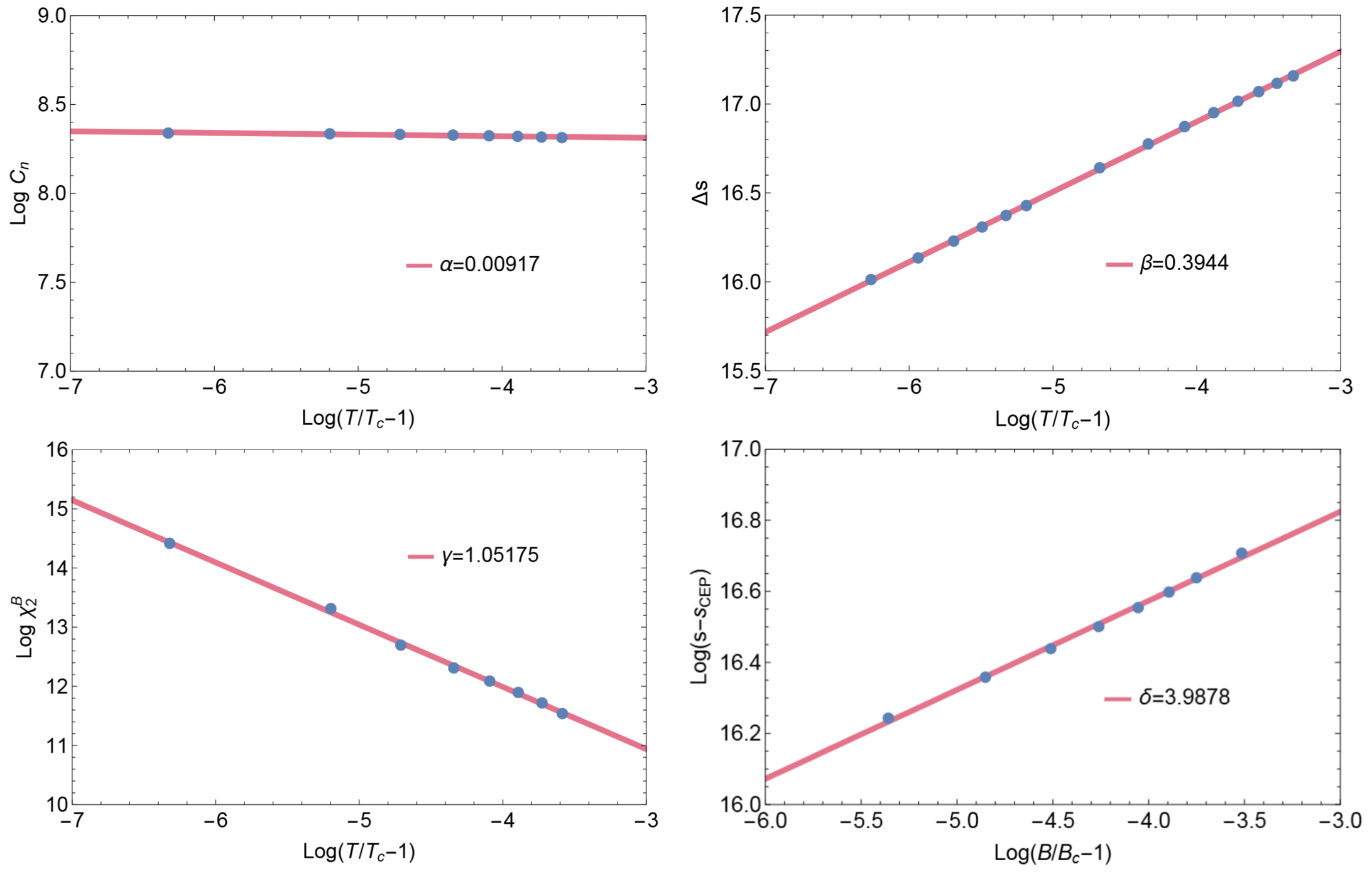}
	\caption{Critical phenomena of the CEP in the $B$-$T$ plane at $\mu_B=0$. \textbf{Top left:} Specific heat $C_n$ along the tangent of the first-order line, yielding the critical exponent $\alpha=0.00917$. \textbf{Top right:} Discontinuity of the entropy density $s$ across the first-order phase transition line, from which the critical exponent $\beta=0.3944$ is extracted. \textbf{Bottom left:} Baryon susceptibility along the first-order axis, which gives the critical exponent $\gamma=1.05715$. \textbf{Bottom right:} Entropy density $s$ as a function of $\mu_B$ on the critical isotherm $T=T_{\text{CEP}}$, yielding the critical exponent $\delta=3.9878$.}\label{betadelta}
\end{figure*}

For illustration, we now present the critical behavior of the CEP in the $B$-$T$ plane at zero $\mu_B$ (see Fig.~\ref{TB-mu-0}).
For example, the discontinuity in entropy density across the first-order phase transition line is displayed in the top-right panel of Fig.~\ref{betadelta}, while the bottom-right panel shows the trajectory of entropy density $s$ as $\mu_B$
approaches $\mu_{B\text{CEP}}$ on the critical isotherm. The power-law scaling is evident in both cases, yielding the critical exponents $\beta=0.3944$ and $\delta=3.9878$. The values of these two critical exponents deviate clearly from the mean-field predictions. Similarly, we obtain the other critical exponents and $\alpha=0.00917$ and $\gamma=1.05175$ from the left panels of Fig.~\ref{betadelta}.
Table~\ref{tabexp} presents the critical exponents for the CEP at three different magnetic field values, denoted by CEP (I, II, III). Although close to mean-field values, these exponents show significant deviations depending on the CEP location, particularly as $B$ increases. Such deviations highlight the features of our holographic QCD model, which cannot be attributed to large-$N$ effects typical in conventional holographic duality, where mean-field behavior is expected. A similar deviation was observed in the holographic 2-flavor model~\cite{Zhao:2023gur}. The critical exponents match those of the quantum 3D Ising model, further emphasizing the distinct nature of our approach compared to traditional large-$N$ QCD models. The gradual change in critical exponents observed at these CEPs is unusual in standard critical phenomena, where transitions between universality classes are typically abrupt. However, this gradual variation in our model arises from the magnetic field's influence on the system's symmetry and degrees of freedom. In particular, the magnetic field reduces certain degrees of freedom, modifying the system’s overall symmetry, which in turn causes the observed gradual transition in critical behavior.
\begin{table}[ht!]
    \setlength{\tabcolsep}{0.5 mm}{
    \begin{tabular}{|c|c|c|c|c|}
     \cline{1-5}
      &    $\alpha$    &   $\beta$        & $\gamma$ & $\delta$ \\ \cline{1-5}
    Experiment & 0.110-0.116 &     0.316-0.327     &   1.23-1.25      &   4.6-4.9        \\ \cline{1-5}
    3D Ising  & 0.110(5) &  0.325$\pm$0.0015 & 1.2405$\pm$0.0015       & 4.82(4)    \\ \cline{1-5}
    Mean field  &  0  &  1/2 & 1          & 3          \\ \cline{1-5} 
    DGR model  &  0   & 0.482          & 0.942          & 3.035   \\ \cline{1-5} 
    CEP I   &  0.002296  & 0.485518          &  0.955819        & 3.00993  \\ \cline{1-5} 
   CEP II  &   0.001694  & 0.50373         &  0.91803        & 2.9455  \\ \cline{1-5}
    CEP III   & 0.00917 & 0.3944          &  1.05175       & 3.9878   \\ \cline{1-5}
    \end{tabular}}
\caption{Critical exponents from experiments in non-QCD fluids, the full quantum 3D Ising model, mean-field (van der Waals) theory, the DGR model~\cite{DeWolfe:2010he}, and our (2+1)-flavor hQCD model. The CEP (I, II, III) correspond to the critical exponents for $\mu_B =554.66\ \text{MeV}, B = 0$ (CEP I), $\mu_B =501.4 \ \text{MeV}, B = 0.3 \ \text{GeV}^2$ (CEP II), and $\mu_B = 0, B = 1.6 \ \text{GeV}^2$ (CEP III), respectively.}
    \label{tabexp}
\end{table}

Before ending this section, we point out that the four critical exponents $(\alpha, \beta, \gamma, \delta)$ are not independent by assuming that the singular part of the free energy density near a CEP is a generalized homogeneous function, from which one obtains two scaling relations:
\begin{equation}\label{scalinglaw}
\alpha + 2\beta + \gamma = 2,\quad  
\delta-\gamma/\beta=1\,.
\end{equation}
The first is the Rushbrooke relation, and the second is the Widom relation.
At a weak magnetic field, within numerical uncertainty, the critical exponents we obtain are quantitatively consistent with both scaling relations, as seen for CEP I and CEP II in Table~\ref{tabexp}. However, a clear violation of the scaling relations emerges at a strong magnetic field. For CEP III in Table~\ref{tabexp} (see also Fig.~\ref{betadelta}), we find that
\begin{equation}
\alpha + 2\beta + \gamma = 1.8498,\quad \delta-\gamma/\beta=1.3212 \,, \nonumber  
\end{equation}
deviating from the conventional scaling laws in~\eqref{scalinglaw}.
These departures can be attributed to several factors. First, numerical uncertainties in the model's calculations can cause small deviations in the exponents. Second, the presence of a strong magnetic field may affect the scaling behavior, leading to departures from the traditional relations. Finally, the violation may reflect intrinsic properties of the holographic model, where quantum fluctuations and the competition among multiple degrees of freedom in strongly correlated systems could break the standard scaling hypothesis. Although violations of~\eqref{scalinglaw} are uncommon, similar deviations have been reported in other black hole systems, see \emph{e.g.}~\cite{Frassino:2014pha,Hu:2024ldp}.

\section{VI. Conclusion}
We have developed a novel neural ODE framework that solves the inverse problem of constructing a holographic QCD action from Lattice QCD data. This framework results in the first holographic model capable of capturing key thermodynamic behaviors of hot and dense QCD at finite magnetic fields. Notably, the model reveals a rich phase structure in a strong magnetic field, including non-monotonic CEP temperature behavior (Fig.~\ref{fig:lattice_comparison1}) and multiple CEPs in the $T$-$\mu_B$ plane (Fig.~\ref{muT-twoCEP}), providing specific experimental markers for validation at future facilities like FAIR, JPARC-HI, and NICA~\cite{Fukushima:2020yzx}. Specifically for $B=1.618 \text{GeV}^2$, we identified the first CEP at ($T_C = 87.3 , \mathrm{MeV}, \mu_C = 115.9 , \mathrm{MeV}$), and a second CEP at ($T_C = 78.9 , \mathrm{MeV}, \mu_C = 244.0 , \mathrm{MeV}$). These results suggest a richer phase structure in QCD than previously anticipated. Furthermore, we have determined critical exponents that depend on the location of CEPs, offering valuable insights for experimental studies in regions accessible to RHIC and LHC. Experimental observables, such as baryon number or magnetization fluctuations, could directly test these predictions in current experiments, \emph{e.g.} RHIC~\cite{STAR:2020tga}, the STAR fixed target program (FXT), and future experiments~\cite{Fukushima:2020yzx}.

Our work has established a fruitful connection between high-energy nuclear physics, gravity, and machine learning.
We have discovered previously unexplored structures of the QCD phase diagram, providing a deeper understanding of QCD matter in extreme environments. The present study raises several interesting issues that warrant further investigation. In particular, within the large magnetic field regime, the phase boundary appears to split into disconnected segments. While these segments may correspond to distinct phase transitions, the exact nature of these transitions remains unclear within the current model. This limitation stems from the fact that the holographic framework primarily captures phase transitions at the level of free energy, offering only a macroscopic description of the system. To more precisely determine the nature of these phase transitions and their connection to specific QCD phenomena, additional degrees of freedom will be necessary. Specifically, we plan to incorporate order parameters that can capture the chiral and deconfinement phase transitions, which are central to understanding QCD dynamics. This is an important direction for future work. Moreover, it is important to note that while the electromagnetic current and the baryonic current are not fully independent, as described by the Gell-Mann-Nishijima relation, we simplify their treatment by introducing two independent U(1) gauge fields. This assumption is made to reduce the complexity of the model and is commonly used in holographic models. However, it should be noted that in certain extreme conditions, such as very high magnetic fields, the coupling between the two currents may become significant, and this simplification may need to be revisited for further refinement. Finally, it is important to understand the observed violation of the traditional scaling relations~\eqref{scalinglaw} under strong magnetic fields.

Future research should extend the model to bridge the gaps between theoretical predictions and experimental findings across energy scales. This includes incorporating isospin asymmetry, which is relevant for neutron stars, and rotational effects, critical for understanding dynamics in rapidly spinning neutron stars and heavy-ion collisions. Extending the model to non-equilibrium scenarios could also provide insights into the real-time dynamics of phase transitions, capturing rapid changes in temperature, density, and magnetic field during heavy-ion collisions. Applying our findings to neutron stars and early universe conditions, where understanding the equation of state for strongly interacting QCD matter under varying magnetic fields is crucial, presents another promising direction for future research.


\section{acknowledgments}
We thank Heng-Tong Ding, Matti Järvinen, Elias Kiritsis, Zhibin Li, Xiaofeng Luo, Yi-Bo Yang, Danning Li, Javier Subils, Ling-Xiao Wang, and Yuan-Xu Wang for stimulating discussions. This work is supported in part by the National Natural Science Foundation of China, Grants No.\,12075101, No.\,12525503, No.\,12047569, No.\,12588101, No.\,12235016, No.\,12588101 and No.\,12447101. S.H. would also like to express appreciation for the financial support from the Max Planck Partner Group. We acknowledge the use of the High Performance Cluster at the Institute of Theoretical Physics, Chinese Academy of Sciences.

\textbf{Conflict of Interest}.  The authors declare that they have no conflict of interest.

\appendix

\setcounter{equation}{0}

\renewcommand*{\thetable}{S\arabic{table}}
\renewcommand*{\theequation}{S\arabic{equation}}

\section{Appendix A: Equations of motion and thermodynamics}\label{Equationofmotion}
Substituting the ansatz~\eqref{EOM} into~\eqref{fig:EOM} gives six equations:
\begin{equation}\label{EOM2}
\begin{split}
   &\phi^{\prime \prime}(r)+\left(\frac{f^{\prime}(r)}{f(r)}+\frac{1}{2} \left(\frac{6}{r}+\frac{g^{\prime}(r)}{g(r)}-\eta^{\prime}(r)\right)\right) \phi^{\prime}(r)\\
   &\quad \quad \quad \quad  -\frac{V^{\prime}(\phi)}{f(r)}+\frac{e^{\eta(r)} A^{\prime}(r)^{2} Z^{\prime}(\phi(r))}{2 f(r)}-\frac{B^{2} \hat{Z}^{\prime}(\phi)}{2 r^{4} f(r)}=0\,,\\
   &A^{\prime \prime}(r)+\frac{1}{2} A^{\prime}(r) \left(\frac{6}{r}+\frac{g^{\prime}(r)}{g(r)}+\eta^{\prime}(r)+\frac{2 Z^{\prime}(\phi) \phi^{\prime}(r)}{Z(\phi)}\right)=0\,,\\
   &\eta^{\prime}(r)
   +\frac{-r^3 (f(r)+r f^{\prime}(r)) g^{\prime}(r)}{r^3 f(r) \left(3 g(r)+r g^{\prime}(r)\right)}\\
   &\quad \quad   \quad  \quad \quad \quad \quad \quad     +\frac{g(r) \left(B^2 \hat{Z}(\phi)+r^4 f(r) \phi^{\prime}(r)^2\right)}{r^3 f(r) \left(3 g(r)+r g^{\prime}(r)\right)}=0\,,\\
   &f^{\prime}(r)+\frac{1}{2} f(r) \left(\frac{4}{r}+\frac{g^{\prime}(r)}{g(r)}-\eta^{\prime}(r)\right)+\frac{1}{3} r V(\phi)+\frac{B^2 \hat{Z}(\phi)}{3 r^3}\\
   &\quad \quad \quad \quad \quad \quad \quad \quad  \quad \quad \quad \quad   \quad     +\frac{1}{6} e^{\eta(r)} r \hat{Z}(\phi) A^{\prime}(r)^{2}=0\,,\\
   &        g^{\prime \prime}(r)-\frac{g^{\prime}(r)^{2}}{2 g(r)}+g^{\prime}(r) \left(\frac{2}{r}+\frac{\eta^{\prime}(r)}{2}\right)\\
   & \quad \quad  \quad \quad \quad \quad \quad  \quad \quad \quad \quad       +g(r) \left(\frac{3 \eta^{\prime}(r)}{r}+\phi^{\prime}(r)^{2}\right)=0\,,
\end{split}
\end{equation}
where five of them are independent.

Expansion at the UV boundary $r\rightarrow\infty$ yields
\begin{equation}\begin{aligned}\label{UVexpand}
&\phi(r)=  \frac{\phi_s}{r}+\frac{(\phi_v+\frac{1}{6} (-1+6 c_1^4) \phi_s^3 \ln [r])}{r^3}+\frac{B^2 \hat{Z}^{\prime}(0)}{6 r^4}+\cdots,\\
&g(r)=  1+\frac{g_4-\frac{1}{4} B^2 \ln [r] \hat{Z}(0)}{r^4}+\frac{B^2 \phi_s \hat{Z}^{\prime}(0)}{5 r^5}+\cdots,\\
&\eta(r)= 0+\frac{\phi_s^2}{6 r^2}+{g_4\over r^4}+\frac{ \left((1-6 c_1^4) \phi_s^4+72 \phi_s \phi_v+12 B^2 \hat{Z}(0)\right)}{144 r^4}\\
&\quad \quad \quad\quad\quad+\frac{1}{12} \frac{\ln [r] \left(-((1-6 c_1^4) \phi_s^4)-3 B^2 \hat{Z}(0)\right)}{r^4}\\
&\quad\quad\quad\quad\quad\quad\quad\quad\quad\quad\quad\quad+\frac{16 B^2 \phi_s \hat{Z}^{\prime}(0)}{45 r^5}+\cdots,\\
&A(r)= \mu_B-\frac{ 2\kappa_N^2 n_B}{2 r^2}+\frac{ 2\kappa_N^2 n_B \phi_s Z^{\prime}(0)}{3 r^3 Z(0)}\\
&+\frac{2\kappa_N^2 n_B \phi_s^2 \left(Z(0)^2-12 Z^{\prime}(0)^2+6 Z(0) Z^{\prime \prime}(0)\right)}{48 r^4 Z(0)^2}+\cdots,
\end{aligned}\end{equation}
where we have taken the normalization of the spacetime coordinates at the boundary such that $\eta(r\rightarrow\infty)=0$ and $g(r\rightarrow\infty)=1$. Expansion at the event horizon $r=r_h$ gives
\begin{equation}\label{IRexpand}
\begin{aligned}
f = & f_h(r - r_h) + \cdots, \\
\eta = & \eta_h  + \eta_1 (r - r_h) + \cdots, \\
A = & A_h(r - r_h) + \cdots,\\
 \phi = & \phi_h  + \phi_1 (r - r_h) +\cdots,\\
 g = &  g_h + g_1 (r-r_h) + \cdots. 
\end{aligned}
\end{equation}
After substituting~\eqref{IRexpand} into the EoMs~\eqref{EOM2}, one finds five independent coefficients $(r_h, A_h, \eta_h, \phi_h, g_h)$.

The relationship between the free energy density $\Omega$ and the on-shell action $S$ is given by
\begin{equation}
-\Omega V=T (S+S_\partial)_{on-shell}\,,
\end{equation}
where $V$ is the spatial volume of the boundary system. The boundary term is  given by
\begin{equation}
\begin{aligned}\label{UVexpand}
S_\partial=&\frac{1}{2\kappa_N^2} \int_{r \rightarrow \infty} d^{4}x \sqrt{-h} \left[2 K-6-\frac{1}{2} \phi(r)^2\right.\\
&\quad\quad -\left(\frac{6 c_1^4-1}{12}\right) \phi(r)^4 \ln [r]
-b \phi(r)^4\\
&\quad\quad\quad\left. +\frac{1}{4} (F_{\mu \nu} F^{\mu \nu}+\hat{Z}(0) \hat{F}_{\mu \nu} \hat{F}^{\mu \nu}) \ln[r] \right]\,.
\end{aligned}
\end{equation}
Here, $h_{\mu\nu}$ is the induced metric at the UV boundary with $K_{\mu\nu}$ the extrinsic curvature defined by the outward pointing normal vector to the boundary.

The boundary energy-momentum tensor reads
\begin{equation}\begin{aligned}
T_{\mu\nu}= &\lim_{r\rightarrow\infty}\frac{2r^2} {\sqrt{-h}}\frac{\delta (S+S_\partial)}{\delta h^{\mu\nu}}\,,\\
  =& \frac{1}{2\kappa_N^2}\lim_{r\rightarrow\infty}r^2\Big{[}2(K h_{\mu\nu}-K_{\mu\nu}-3h_{\mu\nu})\\
  &-(\frac{1}{2}\phi^2
+\frac{6c_1^4-1}{12} \phi^4 \ln[r]+b \phi^4) h_{\mu\nu}\\
& -(F_{\mu\rho} F_{\nu}^{\rho}-\frac{1}{4} h_{\mu\nu} F_{\rho\lambda} F^{\rho\lambda}) \ln[r]\\
&- \hat{Z}(0)(\hat{F}_{\mu\rho} \hat{F}_{\nu}^{\rho}-\frac{1}{4} h_{\mu\nu} \hat{F}_{\rho\lambda} \hat{F}^{\rho\lambda}) \ln[r]\Big{]}\,.
\end{aligned}
\end{equation}
Substituting the UV expansion on the boundary gives:
\begin{equation}\begin{aligned}
&\epsilon=T_{tt}=\frac{(1 + 48 b) \phi_s^4}{96 \kappa_N^2} + \frac{\phi_s \phi_v}{2 \kappa_N^2} \\
&\quad\quad\quad\quad\quad\quad\quad+ \frac{-144 f_v + 192 g_4 + 12 B^2 \hat{Z}[0]}{96 \kappa_N^2}\,,
\end{aligned}
\end{equation}
\begin{equation}\begin{aligned}
&p_x=p_y=T_{xx}=T_{yy}=\frac{(3 - 48 b - 8 c_1^4) \phi_s^4}{96 \kappa_N^2}\\
&\quad\quad\quad\quad\quad\quad\quad + \frac{\phi_s \phi_v}{2 \kappa_N^2} + \frac{-48 f_v - 8 B^2 \hat{Z}[0]}{96 \kappa_N^2}\,,
\end{aligned}
\end{equation}
\begin{equation}\begin{aligned}
&-\Omega=p_z=T_{zz}=\frac{(3 - 48 b - 8 c_1^4) \phi_s^4}{96 \kappa_N^2}\\
&\quad\quad\quad\quad+ \frac{\phi_s \phi_v}{2 \kappa_N^2} + \frac{-48 f_v + 4 (48 g_4 + B^2 \hat{Z}[0])}{96 \kappa_N^2}\,.
\end{aligned}\end{equation}
One finds that in the thermodynamic limit $V\rightarrow \infty$, $\Omega=-p_z$.
Note that, in most studies, \emph{e.g.}~\cite{Critelli:2017oub,Knaute:2017opk,Yang:2020hun,Grefa:2021qvt}, the thermodynamic variables were obtained by integrating the standard first law of thermodynamics, whose validity is still under investigation for AdS black holes with scalar hair~\cite{Hertog:2004ns,Anabalon:2015xvl,Lu:2014maa,Li:2020spf}.

From the EoMs~\eqref{EOM2}, we can get a radially conserved charge:
\begin{equation}\begin{aligned}
&Q=e^{\frac{\eta(r)}{2}} \sqrt{g(r)} r^3 \left(r^2 \left(e^{-\frac{\eta(r)}{2}} \frac{f(r)}{r^2}\right)'-Z(\phi) A(r) A^{\prime}(r)\right)\\
&\quad\quad\quad\quad-B^2 \int_{r_h}^{r} \frac{e^{-\frac{\eta(r_s)}{2}} \sqrt{g(r_s)} \hat{Z}[\phi(r_s)]}{r_s} dr_s\,,
\end{aligned}\end{equation}
which connects data from the horizon to the UV boundary. Evaluating it at both the horizon and the boundary yields
\begin{equation}\begin{aligned}
Q=T s=\epsilon-\Omega-\mu_B n_B=\epsilon^{total}-\Omega-\mu_B n_B-B M\,,
\end{aligned}\end{equation}
where $\epsilon^{total}=\epsilon+\epsilon^{field}$ is the total energy including the external field $\epsilon^{field}=BM$ with $M$ the magnetization. This is the expected thermodynamic relation.
More precisely, $M$ can be computed by the partial derivative of the free energy with respect to $B$.
\begin{equation}\begin{aligned}
&M=-\left(\frac{\partial \Omega}{\partial B}\right)_{T,\mu_B}=-\int_{r_h}^{\infty}\frac{B \sqrt{e^{-\eta(r)} g(r)} \hat{Z}[\phi(r)]}{r} dr\\
&\quad\quad\quad\quad+ \lim_{r \to \infty} \frac{B \sqrt{e^{-\eta(r)} f(r) g(r)} \ln[r] \hat{Z}[0]}{r}\,.
\end{aligned}
\end{equation}
It can be checked straightforwardly that the first law of thermodynamics
\begin{equation}
d\Omega=-s dT-n_B d\mu_B-M dB\,,
\end{equation}
is satisfied. One can then obtain the magnetic susceptibility $\chi_B=\left(\frac{\partial M}{\partial B}\right)_{T,\mu_B}$.

Following~\cite{Cai:2022omk}, we choose $c_1=0.7100, c_2=0.0037, c_{3}=1.935, c_{4}=0.085, c_{5}=30$ of~\eqref{VZapp}. Moreover, we take $\kappa_{N}^{2} =2 \pi({1.68}), \phi_s={1085} \mathrm{MeV}$ and $b=-0.27341$. The comparison of our model to experimental data can be found in~\cite{Li:2023mpv,He:2023ado}.

\section{Appendix B: Calculation method and neural ODEs}\label{neuralnetwork}

Neural networks and neural ODEs have been intensively utilized in hQCD literature (\emph{e.g.}~\cite{Akutagawa:2020yeo, Hashimoto:2020jug, Hashimoto:2022eij, Chen:2024ckb}). In particular, neural networks have been effectively integrated into ODE frameworks in{~\cite{Hashimoto:2020jug, Hashimoto:2022eij}}. In our approach, we introduce a novel neural ODE architecture to numerically solve the magnetic coupling $\hat{Z}[\phi(z)]$, constrained by lattice QCD data with high precision. We model $\hat{Z}$ using a feedforward neural network with three hidden layers, each structured as $x = \sigma(\text{weight} \times x' + \text{bias})$, where the activation function is $\sigma = \tanh$. Here, $x$ and $x'$ represent the output and input of each layer, respectively, and $H = \{\text{weight, bias}\}$ is the parameter set. The layer structure is [input(1)-(16)-(64)-(16)-output(1)]. Details for reproducibility are provided below.
\begin{figure}[hbt!]
     \centering
     \includegraphics[width=0.5\textwidth]{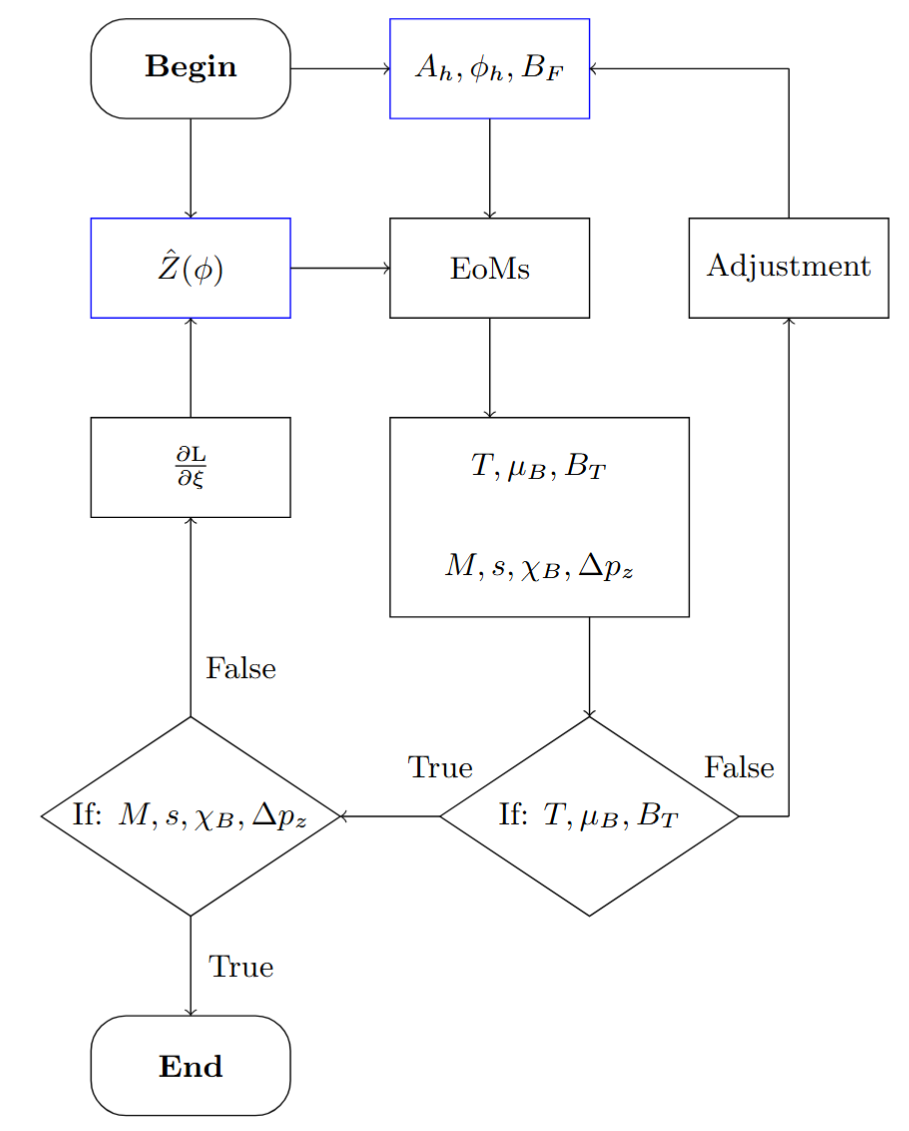}
        \caption{Illustration of the  
     Algorithm process: Given a trial functional \( \hat{Z} (\phi)\) and a set of $(A_h, \phi_h, B_F)$. Solving the EoMs~\eqref{EOM2} to obtain the thermodynamic quantities $(T, \mu_B, B_T, M, S, \chi_B, p_z)$. Verify whether these $(T, \mu_B, B_T)$ cover the range of lattice QCD data. If not, adjust $(A_h, \phi_h, B_F)$. If they do, compare this set with the corresponding lattice data for $(M, S, \chi_B, \Delta p_z)$. If consistent, terminate the process. If not, adjust \(\hat{Z}(\phi)\) and repeat the process. Adjustments to \(\hat{Z}(\phi)\) are made through gradient descent, where \(\text{L}\) represents the loss function and \(\xi\) are the network parameters used to mimic \(\hat{Z}(\phi)\). \(\frac{\partial \text{L}}{\partial \xi}\) indicates the direction of descent for the loss function. When the loss function reaches its minimum, it signifies the optimal solution for \(\hat{Z}(\phi)\).}
     \label{fig: Algorithm process}
 \end{figure}
Fig.~\ref{fig: Algorithm process} illustrates our computational approach. To address the inverse problem of mapping lattice QCD data to a holographic model, we initialize a trial function $\hat{Z}(\phi)$, used to solve the bulk EoMs~\eqref{EOM2} with asymptotic AdS boundary conditions and regular horizon conditions. The solution yields boundary field theory observables, which are then compared to lattice QCD data to iteratively refine $\hat{Z}(\phi)$. Due to scaling symmetries, there are three independent IR data points: $A_h$, $\phi_h$, and $B_F$, corresponding to the field values at the event horizon and a pre-scaled magnetic field. These map to the UV quantities—temperature $T$, chemical potential $\mu_B$, and physical magnetic field $B_T$—where $B_T$ is the transformed field after scaling. By modeling $\hat{Z}(\phi)$ through a neural network, we solve the EOMs and obtain thermodynamic quantities such as $T$, $\mu_B$, $B_T$, $M$, $s$, $\chi_B$, and $\Delta p_z$.

Since the lattice data~\cite{Bali:2014kia} covers only a small region at zero chemical potential, the values of $T$, $\mu_B$, and $B_T$ computed with an arbitrary set of $A_h$, $\phi_h$, and $B_F$ cannot adequately cover this region. Therefore, as shown in Fig.~\ref{fig: Algorithm process}, it is necessary to adjust $A_h$, $\phi_h$, and $B_F$ to obtain a set that effectively covers the relevant lattice QCD region. The values of the remaining thermodynamic quantities $M$, $S$, $\chi_B$, and $\Delta p_z$ depend on the choice of the trial function $\hat{Z}(\phi)$. To optimize $\hat{Z}(\phi)$, we define a loss function $\text{L}=\text{L}(M, s, \chi_B, \Delta p_z)$. We can obtain the optimal $\hat{Z}(\phi)$ by iteratively applying gradient descent (Adam: $\alpha$ =0.0002, $\beta_1$=0.9, $\beta_2$=0.999) to minimize the loss function \cite{KingmaAndBa2014}. Since our model's high precision requirements, we have to employ a neural network ODE model~\cite{chen2018neural} to solve for $\hat{Z}(\phi)$ throughout the entire process. This model effectively transforms the conventional neural network into a continuous form, facilitating the differential equations' rapid and accurate solutions. 

\begin{figure}[]
    \includegraphics[width=0.5\textwidth]{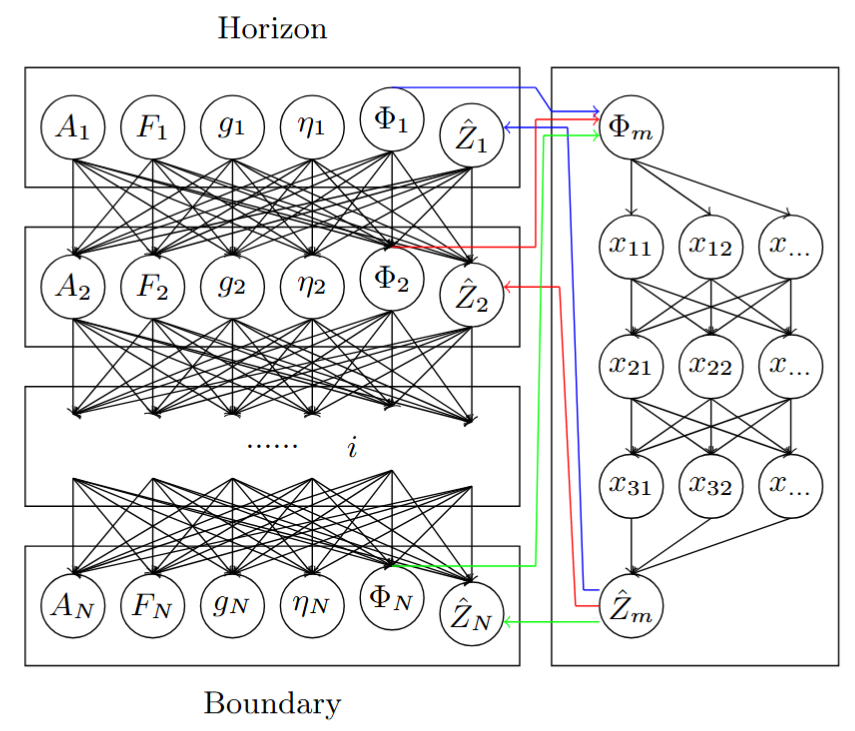}
        \caption{Discrete network representation of the recursive relationship between layers in solving (\ref{equations1st}). Initial conditions are set at the event horizon ($i = 1$), and layers extend to the UV boundary ($i = N$). Each layer $i$ corresponds to field values $ \Theta_i $. The network discretizes the holographic direction $z$ into steps $dz$. $\hat{Z}(\Phi)$ is modeled by a feed-forward network with three hidden layers. $\{x_{11},x_{12},\dots \}$ is outputs of the hidden layer. The colored arrows (\emph{e.g.} green, blue, red) indicate different layers sharing the same functional form of $\hat{Z}(\Phi)$. The arrow is the forward propagation direction, and the opposite is the backpropagation direction.}
        \label{fig:NN on Holo}
\end{figure}
For later convenience, one can rephrase the EoMs~\eqref{EOM2} as the following form:
\begin{equation}
    \begin{aligned}
\frac{d \Theta}{dz}=\Xi(z,\Theta, \dot{\Theta}(z), \hat{Z}(\Phi), {\hat{Z}^{\prime}(\Phi)},B_F),\Theta(z)=\left(\begin{array}{c}
    \Phi (z)\\
    F(z) \\
    \eta(z) \\
    A(z) \\
    g(z)
    \end{array}\right)\,,
\end{aligned}
\label{equations1st}
\end{equation}
where $\dot{\Theta}(z)$ is to take the derivative with respect to the argument and $z=1/r,z \Phi(z)=\phi({ r}), F(z)=z^2 f({ r}), A(z)=A_t({r})$, {$\hat{Z}^{\prime}(\Phi)$ is the derivative with respect to $\Phi$}. One can refer to the precise definitions of these functions of~\eqref{EOM2}. $\Xi$ is a five-component vector. $\Theta$ contains scalar field $\phi$, metric components $f$, $g$, $\eta$, and Maxwell field $A_t$. These equations of motion (\ref{equations1st}) can be rewritten as a discrete difference equation:
\begin{equation}    \Theta_{i+1}=\Theta_{i}+\Xi(z_{i},\Theta_{i}, \dot{\Theta_{i}}(z), \hat{Z}(\Phi_{i}),{\hat{Z}^{\prime}(\Phi_{i})},B_F)dz
    \label {eq:i to i+1}\,,
\end{equation}
where we discretize the holographic direction $z$ with a step size $dz$. The index $i$ corresponds to the $i$-th layer. The equation gives the recursive relationship between the $i$-th layer and the $i+1$-th layer. $\Theta_{i}$ corresponds to the field value at the $i$-th layer. Here, $i=1$ represents the event horizon, and $i=N$ corresponds to the UV boundary. As shown in Fig.~\ref{fig:NN on Holo}, the difference equation can be naturally understood as a $6\times N $ network without an activation function.

Here, we select four thermodynamically independent data sets $\textit{\textbf{S}}$ that contain quantities $\textit{\textbf{S}}=\{M, s, \chi_B, \Delta p_z\}$ for an accurate comparison between the holographic model and Lattice QCD data, performing a global fitting. The key problem is to minimize the loss function by optimizing the functional $\hat{Z}(\phi)$. To determine the optimizing direction of $\hat{Z}(\phi)$, one needs backpropagation of the neural network to extract the data associated with ${\partial \text{L}\over \partial \xi}$. $\xi \in H$ is any parameters in $\hat{Z}$ neural networks. And, we choose loss function $\text{L}=\text{L}(M, s, \chi_B, \Delta p_z)$ as mean-square error (MSE). Here, we apply a similar definition of the loss function offered by~\cite{Shi:2022yqw}. The precise form of loss function $\text{L}$ reads 
\begin{align}
    \text{L}&=\sum_{I \in \textit{\textbf{S}}}P_I (I_{\text{LQCD}}-I_{\text{HQCD}})^2 \,,
\end{align}
where $I_{\text{LQCD}}, I_{\text{HQCD}}$ correspond to the thermal dynamical quantities $\textit{\textbf{S}}$ of lattice QCD data and are predicted by holographic QCD, respectively. The $P_I$ is the inverse of the uncertainty, the maximum difference between the LQCD data and its central value.

We must input $\partial \text{L}\over \partial \xi$ and $\xi$ into Adam to minimize the loss function. 
The key issue is to collect $\partial \text{L}\over \partial \xi$. 
For the $i$-th layer, we have the following chain rule:
\begin{equation}
\frac{\partial \text{L}}{\partial \xi_{i}}=\frac{\partial \text{L}}{\partial \Theta_{i+1}}\frac{\partial \Theta_{i+1}}{\partial \xi_{i}}\,.
\end{equation}
Here, $\xi_i$ is the $\xi$ of the $i$-th layer, $\Theta_i$ is the $\Theta$ of the $i$-th layer. One has to note $\xi_i=\xi_j, i\neq j$ that means $\xi_i$ in each layer are the same, but $\frac{\partial \text{L}}{\partial \xi_{i}} \neq \frac{\partial \text{L}}{\partial \xi_{j}}$. From (\ref{eq:i to i+1}), at each layer it can be expressed by 
\begin{equation}
\frac{\partial \text{L}}{\partial \xi_{i}}=\frac{\partial \text{L}}{\partial \Theta_{i+1}}\frac{\partial \Xi(z_{i},\Theta_{i},\hat{Z}(\Phi_{i}),{\hat{Z}^{\prime}(\Phi_{i})},B_F)}{\partial \xi_{i}}dz\,.
\end{equation}
Finally, for the whole network, the key ingredient $\frac{\partial \text{L}}{\partial \xi_{i}}$ is the sum of all partial derivatives:
\begin{equation}\label{Fdata}
    \frac{d \text{L}}{d \xi} =\int \frac{\partial \text{L}}{\partial \Theta}\frac{\partial \Xi}{\partial \xi}dz \nonumber=\int \frac{\partial \text{L}}{\partial \Theta} ( \frac{\partial \Xi}{\partial \hat{Z}} \frac{\partial \hat{Z}}{\partial \xi}+\frac{\partial \Xi}{\partial \hat{Z}^{\prime}} \frac{\partial \hat{Z}^{\prime}}{\partial \xi} )dz\,. \end{equation}
To obtain the first factor of the integrant in (\ref{Fdata}), we can make use of the following chain rule for the two neighborhood layers:
\begin{equation}
\frac{\partial \text{L}}{\partial \Theta_{i}}=\frac{\partial \text{L}}{\partial \Theta_{i+1}}\frac{\partial \Theta_{i+1}}{\partial \Theta_{i}}\,,
\end{equation}
where \(\frac{\partial \text{L}}{\partial \Theta_{i}}\) represents the derivative of each component in \(\Theta\) at the \(i\)-th layer, with the component index omitted for clarity. Here, \(\frac{\partial \Theta_{i+1}}{\partial \Theta_{i}}\) is a \(5 \times 5\) matrix. From (\ref{eq:i to i+1}), we obtain:
\begin{equation}
    \begin{aligned}
        &\frac{\partial \text{L}}{\partial \Theta_{i}}=\frac{\partial \text{L}}{\partial \Theta_{i+1}}\frac{\partial \Theta_{i+1}}{\partial \Theta_{i}}\\
&=\frac{\partial \text{L}}{\partial \Theta_{i+1}}(1+\frac{\partial \Xi(z_{i},\Theta_{i}, \dot{\Theta_{i}}(z),\hat{Z}(\Phi_{i}),{\hat{Z}^{\prime}(\Phi_{i})},B_F)}{\partial \Theta_{i}}dz)\,.\label{eqYY}
    \end{aligned}
\end{equation}

For convenience, let $ y_i $ denote $\frac{\partial \text{L}}{\partial \Theta_{i}}$, and $y$ denote $\frac{\partial \text{L}}{\partial \Theta}$. Then, the above equation can be written as:
\begin{equation}
y_{i}=y_{i+1}\left(1+\frac{\partial \Xi(z_{i},\Theta_{i}, \dot{\Theta_{i}}, \hat{Z}(\Phi_{i}), {\hat{Z}^{\prime}(\Phi_i)},B_F)}{\partial \Theta_{i}}dz\right)\,,
\end{equation}
which corresponds to the following differential form:
\begin{equation}
    y'(z)=-y(z)\frac{\partial \Xi(z,\Theta, \dot{\Theta}(z), \hat{Z}(\Phi),{\hat{Z}^{\prime}(\Phi)},B_F)}{\partial \Theta}\,.
\label{eq:y'}
\end{equation}
To simplify our notation, we note that this set of equations involves five unknown functions as shown in (\ref{equations1st}), and $\frac{\partial \Xi}{\partial \Theta}$ is a $5 \times 5$ matrix.

\begin{figure}[h!]
    \centering
	\includegraphics[width=0.5\textwidth]{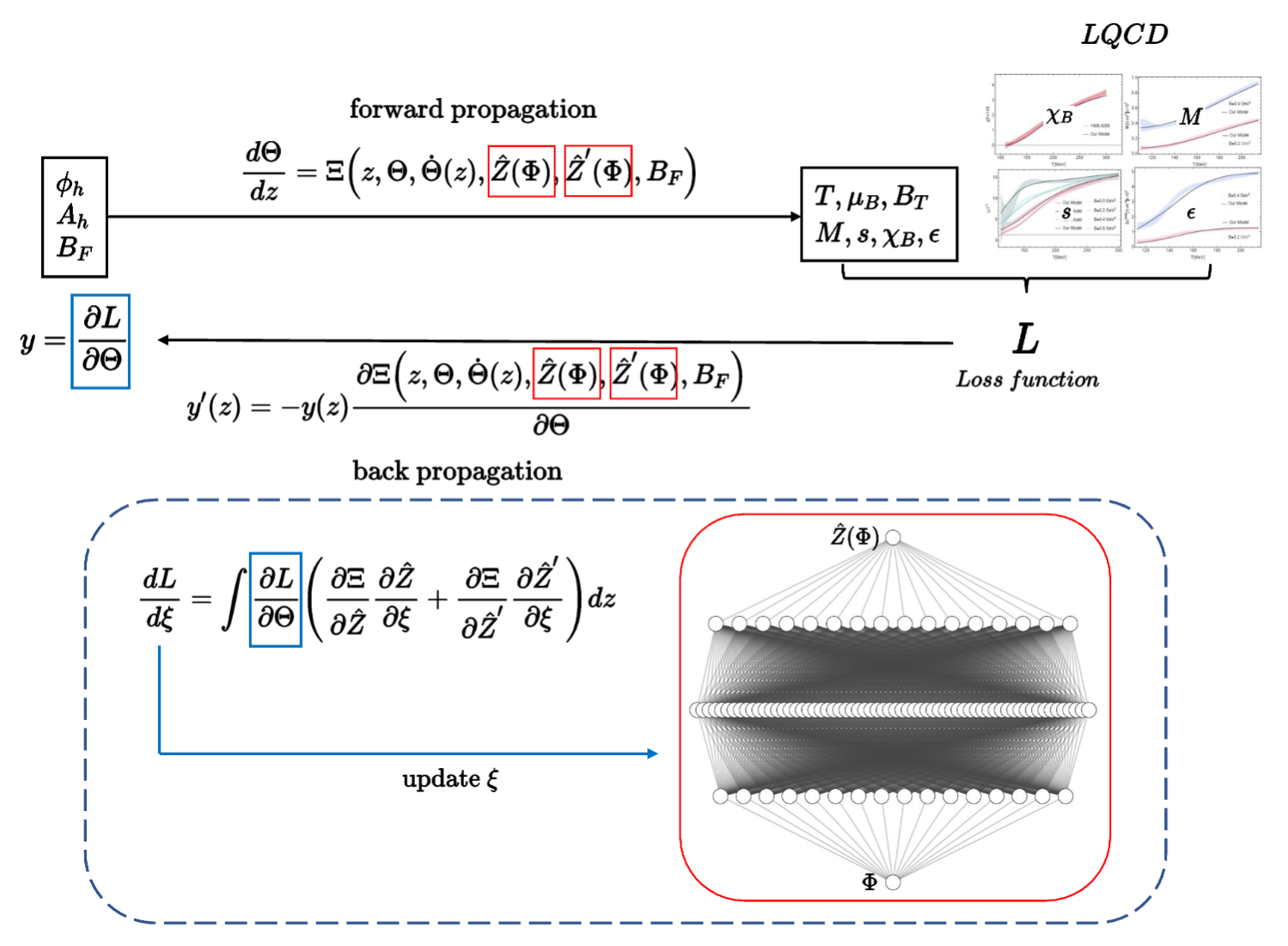}
    \caption{ 
       Continuous representation of our numerical simulation. The top panel depicts the forward propagation, integrating from the infrared (IR, left) to the ultraviolet (UV, right), and comparing with lattice data to compute the loss function. This continuous approach corresponds to Fig.~6. During forward propagation, the function $\hat{Z}$ acts as a numerical component within the equations of motion (EOM). The bottom panel (excluding the dashed section) illustrates backpropagation in the continuous limit. By deriving backward integral equations from the forward ones, we calculate the derivative of the loss function $\text{L}$ with respect to the parameters $\xi$, subsequently updating the parameters in the $\hat{Z}(\phi)$ neural network (dashed section). Specifically, $\frac{\partial \hat{Z}}{\partial \xi}$ and $\frac{\partial \hat{Z}'}{\partial \xi}$ are obtained through the internal back propagation of $\hat{Z}$ across the entire integration domain.
}
    \label{fig:enter-label}
\end{figure}

We elaborate on the forward and backpropagation for the discrete case and derive the continuous form used in practical computations. In the actual calculation, as shown in Fig.~\ref{fig:enter-label}, we first perform a forward propagation integral to obtain the loss function, and then the derivative of the loss function with respect to the parameters can be propagated through the back differential equation, which used for optimizing the trial function $\hat{Z}(\Phi)$ via gradient descent to minimize the loss function $\text{L}$. The forward and backpropagation of the $\hat{Z}$ neural network are respectively regarded as numerical functions participating in the forward and back differential equations.

Finally, we combine all the elements in (\ref{Fdata}) and input them into the Adam optimizer to achieve the functional $\hat{Z}(\phi)$, which is a crucial point of this work. We managed to obtain the numerical data for $\hat{Z}(\phi)$ as shown in Fig.~\ref{fig:AIZphi}. It can be good approximated using the following analytical form:
\begin{equation}\label{myfit}
    \begin{aligned}
    \hat{Z}(\phi) &= a_0 e^{-a_1 (\phi - a_2)^2} + a_3 e^{-a_4 (\phi - a_5)^2 - a_6 (\phi - a_7)^4}\\&
    +a_8 \mathrm{sech}[-a_9 (\phi - a_{10})^2] + a_{11} e^{-a_{12} (\phi - a_{13})^6} + a_{14}\,,
    \end{aligned}
\end{equation}
where the parameters are given by
\begin{equation}
\begin{aligned}
a_0 &= \frac{49677}{100000}, & a_1 &= \frac{8583}{25000}, & a_2 &= \frac{202953}{100000}, \\
a_3 &= \frac{15371}{50000}, & a_4 &= \frac{6297}{50000}, & a_5 &= \frac{39131}{20000}, \\
a_6 &= \frac{411}{50000}, & a_7 &= \frac{413981}{100000}, & a_8 &= \frac{97}{4000}, \\
a_9 &= \frac{34873}{100000}, & a_{10} &= \frac{29503}{50000}, & a_{11} &= -\frac{287}{50000}, \\
a_{12} &= \frac{24319}{12500}, & a_{13} &= \frac{2637}{2500}, & a_{14} &= -\frac{691}{50000}.
\end{aligned}
\end{equation}

\begin{figure}[ht!]
	\centering
    \includegraphics[width=0.43\textwidth]{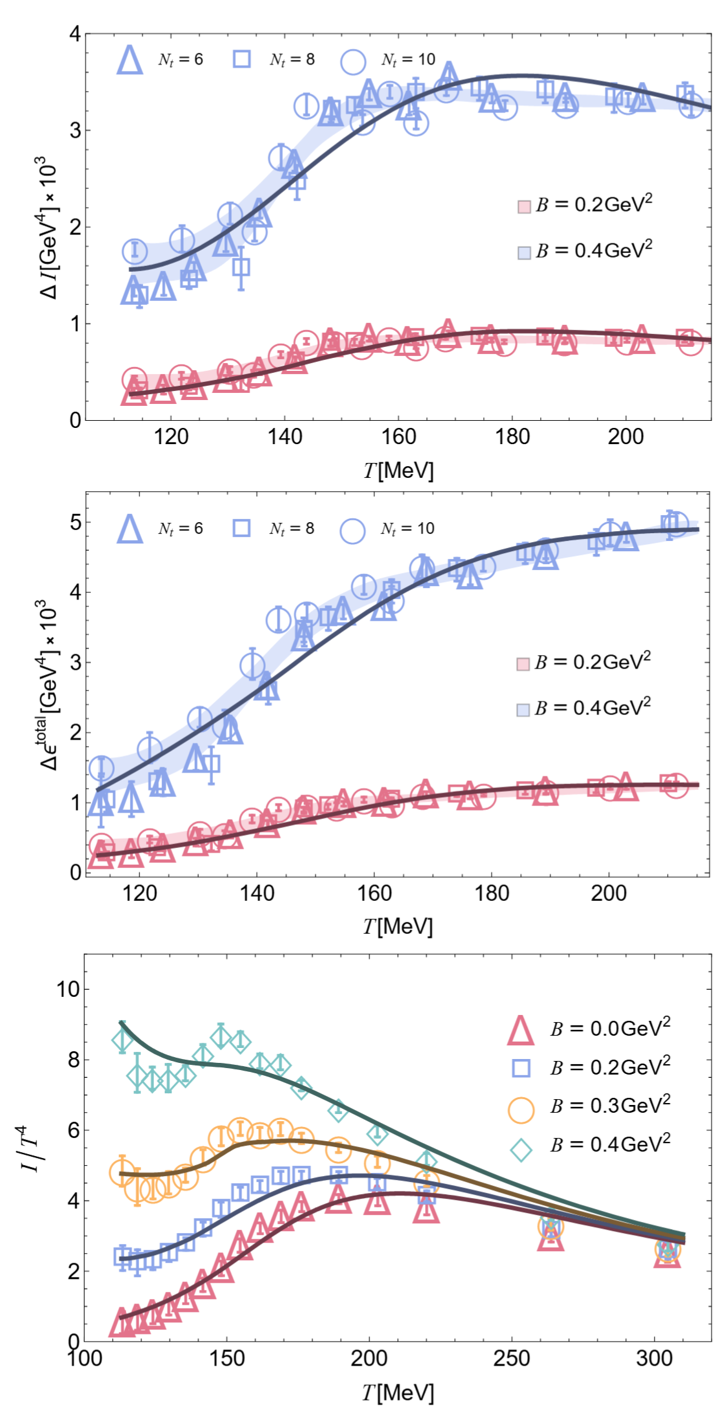}
	\caption{The renormalized trace anomaly $\Delta I$ (top), the renormalized energy density $\Delta \epsilon=\epsilon|_{B}-\epsilon|_{B=0}$(middle) and the trace anomaly $I$(bottom). Our holographic computations (solid curves) are compared with the latest lattice QCD results from~\cite{Bali:2014kia}. The $N_t$ corresponds to three lattice spacings, and $B$ denotes the magnetic field strength. The shaded areas correspond to lattice continuum estimates.}
	\label{IPdeltaI}
\end{figure}
To illustrate the efficacy of the algorithm, we present a comparison of four thermodynamically independent quantities, $\textit{\textbf{S}} = \{M, s, \chi_B, \Delta p_z\}$, between the holographic predictions and the lattice QCD simulations, as shown in Fig.1 of the main text. Additionally, we confirm that the corresponding trace anomaly $I$, the renormalized longitudinal pressure $\Delta p_z$, and the renormalized anomaly $\Delta I$ predicted by the holographic model align with the lattice QCD data~\cite{Bali:2014kia}, as depicted in Fig.~\ref{IPdeltaI}. This work represents the first quantitative realization of state-of-the-art lattice QCD data~\cite{Bali:2014kia} within a holographic model.

\end{document}